\documentclass[twocolumn]{revtex4}
\usepackage{graphicx}
\usepackage[]{epsfig}

\newcommand{\beq}{\begin{equation}}
\newcommand{\eeq}{\end{equation}}
\newcommand{\beqd}{\begin{displaymath}}
\newcommand{\eeqd}{\end{displaymath}}
\newcommand{\beqa}{\begin{eqnarray}}
\newcommand{\eeqa}{\end{eqnarray}}

\newcommand{\comment}[1]{}

\newcommand{\bew}{\begin{widetext}}
\newcommand{\eew}{\end{widetext}}

\begin{document}

\title{Fate of the Hybrid Transition of Bootstrap Percolation in Physical Dimension}

\author{Tommaso Rizzo$^{1,2}$}
\affiliation{
$^1$ ISC-CNR, UOS Rome, Universit\`a "Sapienza", PIazzale A. Moro 2, I-00185, Rome, Italy.
$^2$ Dip. Fisica, Universit\`a "Sapienza", Piazzale A. Moro 2, I-00185, Rome, Italy}

\begin{abstract}
Bootstrap, or $k$-core, percolation displays on the Bethe lattice a mixed first/second order phase transition with both a discontinuous order parameter and diverging critical fluctuations.
I apply the recently introduced $M$-layer technique to study corrections to mean-field theory showing that at all orders in the loop expansion the problem is equivalent to a spinodal with quenched disorder. This implies that the mean-field hybrid transition does not survive in physical dimension. Nevertheless, its critical properties as an avoided transition, make it a proxy of the avoided Mode-Coupling-Theory critical point of supercooled liquids.
\end{abstract}

\maketitle

Bootstrap, or $k$-core, percolation (BP) is an extension of ordinary percolation in which the occupancy of a given site is constrained by that of its neighbors.  The definition of the problem is very simple: first the sites of a given
lattice are populated with probability $p$, then each site
with less than $k$ neighbors is removed and the process is repeated until each site has at least $k$ neighbors.
When the process is completed the remaining occupied sites, if any, form the so-called the $k$-core. 
It was originally introduced as a model of dilute magnetic materials but has then been invoked in many different contexts, see \cite{Degregorio2016} for a review. 
In its mean-field (MF) Bethe-lattice formulation \cite{chalupa1979bootstrap} it displays a mixed first/second order phase transition characterized by a discontinuous order parameter and a diverging critical fluctuations.
The presence of a hybrid transition has later been confirmed  in many other locally tree-like random graphs, including random networks with generic connectivity \cite{PhysRevE.73.056101} and heterogeneous constraints \cite{PhysRevLett.107.175703,PhysRevE.87.022134}, while the large connectivity limit has been recently studied in \cite{parisi2015large}. Correspondingly, it has been argued that the culling process by which the $k$-core is generated is also critical at the hybrid transition, although here the exact analysis is difficult also on locally tree-like lattices  and one may need to resort to  numerics \cite{iwata2009dynamics,PhysRevX.5.031017,PhysRevE.94.062307}.
The hybrid nature of the BP transition itself has been invoked in a wide range of contexts \cite{Degregorio2016}, {\it e.g.} network collapse \cite{PhysRevLett.109.248701}, jamming \cite{schwarz2006onset,morone2018jamming} and  Kinetically Constrained Models (KCM) of supercooled liquids \cite{sausset2010bootstrap,PhysRevLett.115.225701,de2016scaling}.
Since most of these applications are  defined in three dimensions an essential question arises: does  the hybrid MF transition exist in finite dimension?  

So far exact results on specific lattices in finite dimension $d$ have always provided a negative answer \cite{Degregorio2016}. It is known rigorously that there is no transition in regular lattices (RL) with $k>d$: the $k$-core is empty as soon as $p<1$ \cite{schonmann1992behavior}. Furthermore numerical investigations suggest that on RL with $k=d=4$ the transition is simply first-order with $p_c<1$ \cite{parisi2008k} while for $k=d=3$ is continuous \cite{Degregorio2016,parisi2008k}.
Note that numerical methods are affected by severe non-trivial finite-size effects for $k>d$ \cite{holroyd2003sharp} but are safer for $k \leq d$ \footnote{The essential difference between the two cases is  that the $k$-core for $k\leq d$ may exist in an isolated system with empty boundaries \cite{parisi2008k}}. 
Nevertheless these negative results leave the question open for the countless other lattices (either regular or random) in any dimension. 
Furthermore a  perturbative expansion around the large dimension limit of \cite{harris20051} suggested that both the MF transition and its hybrid character survives.

In the following I will address the problem from the Renormalization-Group (RG) perspective. In the RG framework a microscopic model at the critical point lies on the so-called critical surface and will be driven by the RG flow towards a fixed point (FP). Notably all models in the basin of attraction of a given FP have the same critical exponents leading to the celebrated Universality property. 
The focus of the analysis is thus shifted from specific microscopic models to specific ensembles of critical points associated to RG FP's.
Among all possible FP's the MF FP is particularly important: in the typical scenario 
it develops an unstable direction on the critical surface below the upper critical dimension so that a critical point, no matter how close to it, will flow towards another stable FP.
Thus in the RG framework the question of the fate of the MF transition can be given a precise statement: given a point on the critical surface infinitely close to the MF FP, where will the RG flow take it?

Before discussing the answer to this question for BP let us clarify the connection between RG FP's and microscopic models. {\it A priori} any microscopic model  that can be studied with the MF approximation is a  candidate to be in the basin of attraction of the MF FP or, if unstable, of the stable FP connected to it.
In the context of BP this means that any lattice with connectivity $c>2$ is {\it a priori} a  candidate: this includes  all sorts of lattices, {\it e.g.} RL's, the triangular or honeycomb in $d=2$, diamond cubic in $d=3$, plus random lattices in any dimension.
However, to determine if a microscopic model {\it actually} flows to the MF PF we should solve exactly the RG equation, which is typically unfeasible.
Clearly if we know by other means that a model does not have a critical point ({\it e.g.} RL with $k>d$) we know that they will not flow to the MF FP.  
Besides not all candidates models must have the same RG behavior: for instance it is possible that a model flows to a completely different FP, much as BP on the RL with $k=d=3$ flows towards the FP of ordinary percolation.
On the other hand, the fact that a given candidate does not display a MF-like transition is not enough to claim that none of the countless others microscopic models also do not, one could never be sure of not having picked the wrong lattice.
In short, only a FP RG analysis can give us information valid for {\it all} microscopic models that flow towards the MF FP, but it cannot tell us which models do and do not.

The result presented in the following implies that not only the MF FP is unstable below the upper critical dimension $d_u=8$ but that as soon as we move in the unstable direction the hybrid transition is washed out meaning that there is no stable fixed point reachable from the neighborhood of the unstable MF FP: {\it the fate of the MF hybrid transition is to disappear}.
This explains why such a transition has never been observed in finite dimension: no matter which lattice we consider we will never observe neither the hybrid MF transition nor a transition associated to a stable FP reachable from the neighborhood of the MF FP.
At best one can observe pseudo-MF behavior with an ordinary first-order transition occurring slightly before the correlation length divergence, as indeed seems to happen on the regular lattice with $k=d=4$ \cite{parisi2008k}. 
According to what we said before, the RG analysis is intrinsically local and cannot rule out the possibility that on a given specific lattices BP may display other types of transitions  by flowing towards different unknown FP's, however 
the MF hybrid transition that continues to attracts so much interest in the literature is specifically associated to the unique MF FP considered here.

In the modern theory of critical phenomena, to study the MF FP one does not solve explicitly the RG equations. One considers instead a specific microscopic model with a tunable parameter that controls the distance from the MF FP. This is a continuum field theory ({\it e.g.} the $\phi^4$ theory for ferromagnetism) and the tunable parameter is the bare coupling constant. If the coupling constant is zero one sits exactly on the unstable MF FP, but as soon as it is non-zero the RG flow takes the system to the stable FP with different critical exponents. Because of Universality, computing the critical exponents in the field theory amounts to compute them for all models that flow to the same FP. This computation is then performed resumming the diagrammatic loop expansion of the field theory through standard recipes inspired by RG arguments  \cite{Parisi1988,LeBellac1991,Zinn-justin2002,Cardy1996}.
Until now it was not be possible to implement this program for BP because of the lack of a known field-theoretical formulation, at variance with ordinary percolation where the Fortuin-Kastelein map provides the answer.
\begin{figure}
\centering
\includegraphics[scale=.5]{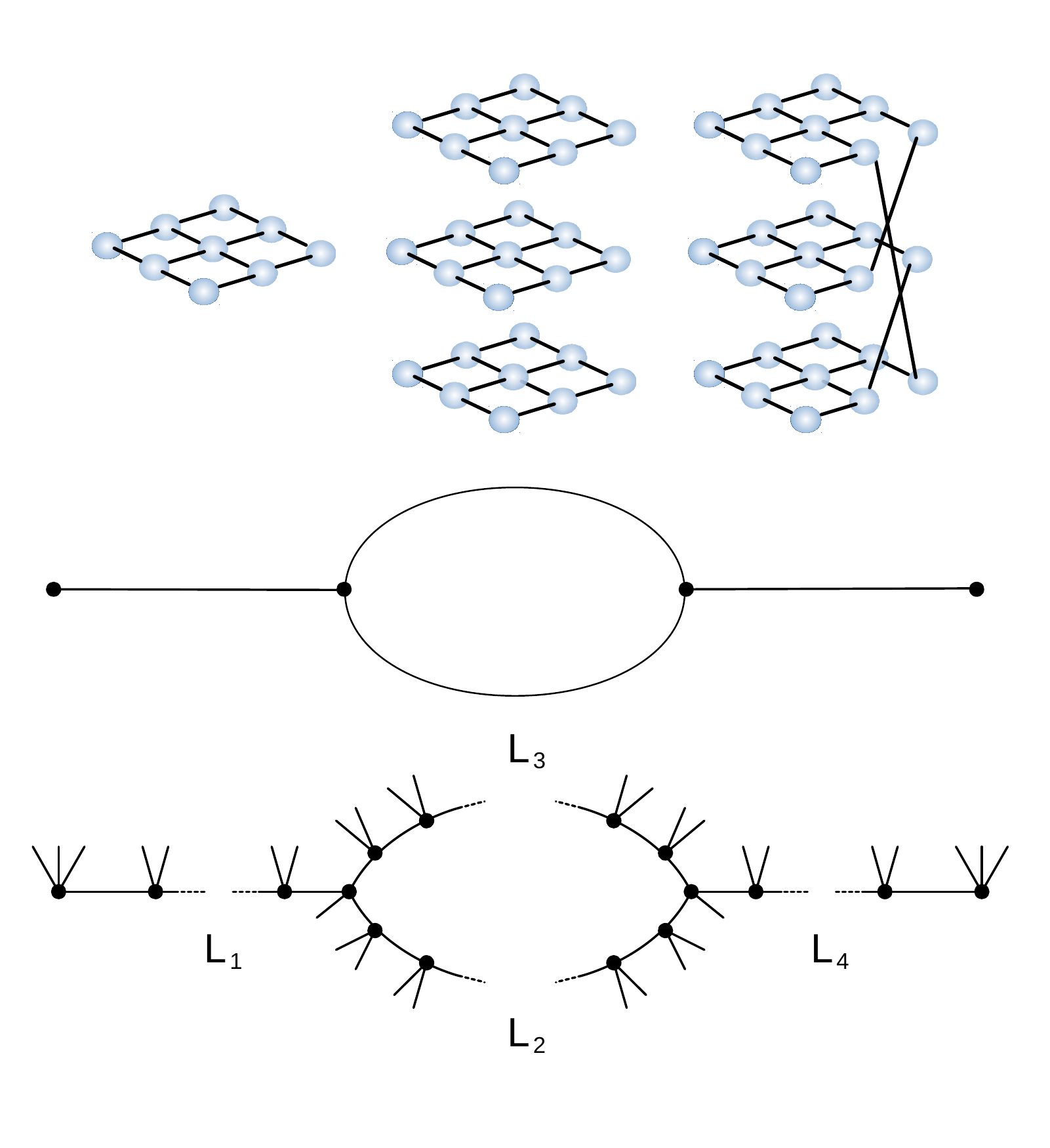} 
 \caption{
The $M$-layer construction: the original lattice is replicated $M$ times and then each edge is rewired between different copies ($M=3$) (top).
The $1/M$ expansion is obtained summing over Feynaman diagrams (middle) that are transformed into fat diagrams (bottom).}
\label{fig:mlayer}
\end{figure}
To overcome this problem I have applied the $M$-layer construction recently proposed in \cite{altieri2017loop} to treat problems in which MF theory is only available on the Bethe lattice. One introduces a microscopic model with an additional parameter $1/M$ that plays the role of the bare coupling constant: when it is zero the Bethe solution is exact and the system flows to the MF FP, when it is small but finite one explores the neighborhood of the MF FP. Furthermore one can compute a loop expansion around the Bethe solution in powers of $1/M$ and, if the MF FP is unstable, compute the non-MF critical exponents by resumming the series through the same recipes of the field theoretical expansions.
The precise technical statement presented in the following is that the loop expansion of BP at all orders is the same of a spinodal with disorder, which is in turn equivalent to a quadratic stochastic equation. The latter is ill-defined beyond perturbation theory and this leads to the fact that the MF FP transition is washed out as soon as one moves in the unstable direction.

The result does not at all curtail the interest and relevance of the MF BP transition for  systems in physical dimension.
Indeed the {\it way} by which the MF transition is washed out, {\it i.e.} the mapping to a spinodal with quenched disorder, consolidates the  status of BP as a proxy of the avoided Mode-Coupling-Theory (MCT) singularity in supercooled liquids.
MCT predicts  a sharp dynamical arrest transition that is not observed in actual supercooled liquids. Although avoided, the transition manifests itself as a crossover whose existence and relevance is well established experimentally.
The connection is established through Fredrikson-Andersen (FA) models of supercooled liquids, that exhibit on the Bethe lattice a sharp MCT transition quantitatively related to BP  \cite{sellitto2005facilitated,PhysRevLett.115.225701,de2016scaling}.
Now this connection is further consolidated because a mapping to the very same quadratic stochastic equations derived here have been obtained in recent years for MCT in the context of the Random-First-Order-Theory \cite{franz2011field} motivating interest in other random critical points \cite{franz2013universality,biroli2014random} and in zero-temperature disordered spinodals \cite{nandi2016spinodals}. 
Later a mapping to a well-defined dynamical stochastic equation, called stochastic beta relaxation (SBR), has been obtained through a full-fledged dynamical treatment \cite{rizzo2014long,rizzo2016dynamical} to provide a complete description of how the sharp transition of MCT is transformed into a dynamical crossover. It is likely that SBR is also valid for FA models, but at present such an analysis is hampered by the lack of a Bethe lattice solution of the dynamics. Indeed at present we are not even able to show analytically that FA models obey critical MCT equations, as shown numerically in \cite{sellitto2005facilitated,PhysRevLett.115.225701,de2016scaling}.
Be as it may, the result of \cite{franz2011field} has already been successfully applied to FA models \cite{franz2013finite,ikeda2017fredrickson}. Similarly, it is natural to expect that the correct dynamical scaling functions in finite dimension are given by SBR as confirmed in recent numerical simulations in the $M$-layer framework  \footnote{T. Rizzo and T. Voigtmann, in preparation.}.

Before sketching the basic steps of the computation I note that the present RG analysis of BP has also a methodological interest in itself because is the first result that, at least at present, can only be obtained through the $M$-layer construction. It is also worth mentioning that some authors have successfully modified {\it ad hoc} the BP rules in order to find a hybrid transition. Initiated by Ref. \cite{Toninelli2008,jeng2008study}, this line of research is rather active \cite{ghosh2014jamming,balister2016subcritical}. These models nonetheless are not directly relevant to the question of the fate of the MF BP transition because: microscopically they are essentially different from BP, they are not amenable to a MF treatment and, most importantly, their hybrid transitions display an essential and striking qualitative difference from MF BP in that the correlation length diverges faster than a power law.

The outcome of the culling process leading to the $k$-core does not depend on the sequence of which  sites are culled, in particular one can consider the  modified problem in which a given  site $i$ is occupied with probability one and no culling is applied to it and then  apply an {\it extraction and culling} (eac) move to obtain its probability $P_{site}$ to be in the $k$-core. The latter is given by $p$ times the sum over all configuration of its neighbors $\{s_j, j \in \partial i\}$ satisfying the constraint of their occupancy probability  $P^{(i)}(s_j, j \in \partial i)$ prior to eac on site $i$.
On the Bethe lattice with connectivity $c$ one can argue that the probability $P^{(i)}(s_j, j \in \partial i)$ is factorized 
\beq
P^{(i)}(\{s_j \, :\,  j \in \partial i\})= \prod_{j \in \partial i}P_B(s_j)
\eeq
where $s=1$ if the site is occupied and zero otherwise and
\beq
P_B(s) \equiv \delta_{s,1}\, P  + \delta_{s,0} (1-P) 
\eeq
With the definition $P_{s,t} \equiv \sum_{i=s}^{t}  {t \choose i} P^i (1-P)^{t-i}$ the problem is determined by the following equations:
\beq
P=p \, P_{k-1,c-1} \, , \ \  P_{site}=p \, P_{k,c} \ . 
\eeq
For all $k>2$ a solution with non-zero $P$ is found at large values of $p$; the solution disappears at a critical value $p=p_c$ with a square-root singularity thus exhibiting the celebrated mixed first-order/second-order character.

To implement the $M$-layer construction we consider $M$ copies of the original lattice (with connectivity $c$) so that on each site $i$ of the original lattice we have $M$ sites $s_i^{\alpha}$, $\alpha=1,\dots, M$. The $M$ edges corresponding to a given edge of the original lattice are then randomly rewired between them  leading to a new lattice with the same connectivity $c$, see fig. (\ref{fig:mlayer}).
One can argue that in the limit $M \rightarrow \infty$ the number of small loops in the graph decreases with $M$ and the new lattice has locally a tree-like structure. Therefore the average over the possible rewirings of any observable takes the value corresponding to the above Bethe lattice solution with small $1/M$ corrections that can be can be expressed as a sum over Feynman diagrams. 

A given Feynman diagram is transformed into a {\it fat diagram} substituting its  $I$ lines with one-dimensional chains of given lengths $\{L_1,\dots,L_I\}$ and attaching the appropriate number of Bethe trees to ensure that the connectivity of each vertex is exactly $c$, see the example of fig. (\ref{fig:mlayer}). The natural observable in the context of BP is the joint probability that two (or more) sites are both on the $k$-core, in the following I will specialize to two-point correlations and thus to Feynman diagrams with two external vertexes but the mapping can be generalized straightforwardly to higher orders correlations.

The computation of the occupation probability on a given fat diagram can be done it two steps: first we study the case in which the vertexes of the Feynman diagram are occupied and perform eac on the internal lines and then we apply eac to the vertexes.
It turns out that in order to evaluate a given diagram we need the probability that sites $s_1$ and $s_{L-1}$ on a chain are occupied prior to eac on the extremal point of the chain $s_0$ and $s_L$; the case in which all sites on the chain are such that they have exactly $m$ occupied neighbors before eac on $s_0$ and $s_L$ must be treated separately, in this case, following standard terminology, we say that they are on the {\it corona} and we introduce an additional binary variable $\chi$ equal to one if this is the case and zero otherwise. 
As discussed in \cite{altieri2017loop}  we have to attach to each line of length $L$ the difference between the probability distribution $P_L(s_1,s_{L-1},\chi)$ and its $L \rightarrow \infty$ limit  $P_B(s_1)P_B(s_L)\delta_{\chi,0}$:
\beq
\Delta P_L(\sigma,\tau,\chi)\equiv P_L(\sigma,\tau,\chi)-P_{B}(\sigma)P_{B}(\tau)\delta_{\chi,0} \ .
\eeq
A careful analysis, presented in the supplemental material, leads to the following large-distance behavior: 
\beqa
\Delta P_L(\sigma,\tau,\chi) & = & \delta_{\chi,0}[g(\sigma)g(\tau)\, \Delta \,L\, \lambda^{L-1}+O(\lambda^L)]+
\nonumber
\\
&+& 
\delta_{\chi,1}\, \delta_{\sigma,1}\, \delta_{\tau,1}\lambda^{L-1}
\label{2point}
\eeqa
where $g(s)\equiv \delta_{s,1}-\delta_{s,0}$  and $\Delta$ is a positive constant. 
The parameter $\lambda$ has the following behavior close to the Bethe critical point: $\lambda \approx (1- \sigma^{1/2})/(c-1)$
where $\sigma$ is a positive quantity vanishing linearly at the Bethe critical point  $\sigma \propto |p-p_c| $. Note that the corona probability decays exactly as $\lambda^{L-1}$ but there is a larger $O(L \, \lambda^L)$ off-corona contribution, this uncommon feature, already noted in \cite{schwarz2006onset}, is the essential to the final result.

According to \cite{altieri2017loop} the Feynman diagram considered must be embedded on the original lattice summing over the positions of the internal vertexes, as in the the standard field theoretical loop expansions. Furthermore for each line of the Feynman diagram we have a sum over all non-backtracking paths connecting the two ends of the line.
In Fourier space the sum of (\ref{2point}) over the non-backtracking paths generates the equivalent of the bare propagator in standard field theory:
\beq
 \delta_{\chi,0}g(\sigma)g(\tau)\,\Delta  \left({1 \over \sigma^{1/2}+  \,k^2 }\right)^2+
\delta_{\chi,1}\, \delta_{\sigma,1}\, \delta_{\tau,1}{1 \over \sigma^{1/2}+ \, k^2 }
\eeq
The key point is that the off-corona contribution leads to a so-called double pole as in systems with quenched random fields. 
The contribution of each Feynman diagram diverges at the Bethe critical point and the modern theory of critical phenomena provides recipes to compute the critical exponents by resumming these divergent contributions. For BP one would say that the leading contribution   of a given Feynman diagram is obtained selecting for each line the double pole contribution. Instead it turns that such a contribution has a zero prefactor, technically because  of the property $\sum_\sigma\,g(\sigma)=0$. A careful analysis (see supplemental material) shows that in order to maximize the number of  off-corona edges  without having a vanishing prefactor one has to arrange the corona edges on trees in such a way that each internal vertex is connected to either $s_1$ and $s_2$ by a path of corona edges. Once this condition is fulfilled we can set the  $\chi$'s of the remaining edges to zero. 

\begin{figure}
\centering
\includegraphics[scale=.5]{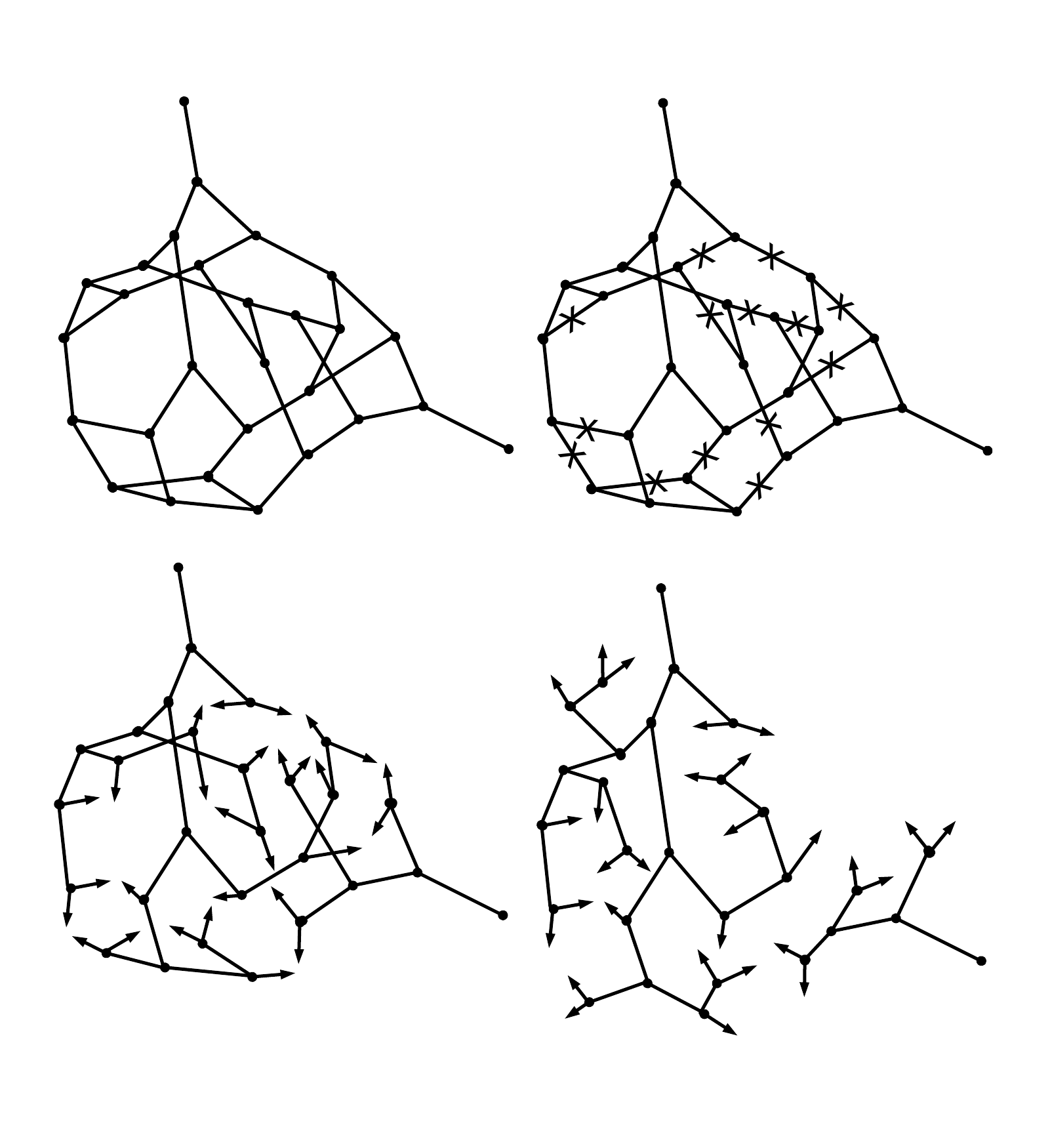} 
 \caption{Top-Left: A cubic 13-loops Feynman diagram contributing to the two-points correlation. Top-Right: one of the possible leading non-vanishing assignments of the subleading $\chi=1$ corona contributions (simple lines) and of the leading $\chi=0$ off-corona contributions (crossed lines).
Bottom: Graphical representation of the mapping between the random field spinodal and quadratic stochastic equations (see text).}
\label{fig:cross}
\end{figure}

In the $M$-layer construction one has to consider all Feynman diagrams with internal vertexes of degree up to  the connectivity of the original lattice and at any given order in the loop expansion the leading divergent diagrams are those where all vertexes have degree three. Therefore the $M$-layer expansion at criticality is naturally associated to cubic Feynman diagrams, unless, which is not the case here,  the contribution of cubic vertexes vanishes for symmetry reason (as for a ferromagnetic transition). 
In fig. (\ref{fig:cross}) we show a leading non-vanishing assignment of the $\chi$'s on a 13-th loop cubic Feynman diagram. We have represented  subleading corona contributions with simple lines and leading off-corona contributions with crossed lines as in the RFIM literature \cite{parisi1992field,de2006random}.

Using standard techniques borrowed from the Ranodm-Field-Ising-Model literature (see the supplemental material for a complete derivation) one can show  that the above Feynman rules for diagrams are exactly those of a magnetic spinodal in a random field.  One can then show that the most diverging diagrams are those generated by the solution of a stochastic equation {\it i.e.} the generic $N$-point correlation function of BP is given by:
\beq
\langle s_{x_1}\dots s_{x_N} \rangle \approx [m(x_1)\dots m(x_N)]
\label{mapping}
\eeq
where $m(x)$ is the solution of the following stochastic equation: 
\beq
{\sigma \over 2 }  +{h(x) \over \tilde{M}}- \nabla^2 m(x)- {1 \over 2}m^2(x)=0
\eeq
and the square brackets mean average over the solution of the  equation with Gaussian random fields $h(x)$
\beq
[h(x)]=0\, , \ \ [h(x)h(y)]= {\Delta} \, \delta(x-y)
\eeq
 The mapping to the stochastic equation is represented graphically in fig. (\ref{fig:cross}). Before average the loop expansion is given by Feynman diagrams with single poles (the corona lines) and random sources (represented by arrows), after average different sources are paired to give the crossed lines (the off-corona lines). 

The average over solutions in the r.h.s. of eq. (\ref{mapping}) is  pathological simply because the equation does not admit a real solution for certain (negative) values of $h(x)$.  The equation is well-defined instead if the random fields are imaginary and yields the universality class of branched polymers \cite{PhysRevLett.46.871}.
As we said before the physical implication is the hybrid critical point ceases to exist as soon as we move in the unstable direction of the MF FP.

\begin{acknowledgments}
I acknowledge the financial support of the Simons Foundation (Grant No. 454949, Giorgio Parisi). This project has received funding from the European Research Council (ERC) under the European Union’s Horizon 2020 research and innovation programme (grant agreement No [694925]).
\end{acknowledgments}

\bibliography{biblio.bib}

\clearpage

\widetext
\begin{center}
\textbf{\large Supplemental Materials: Fate of the Hybrid Transition of Bootstrap Percolation in Physical Dimension}
\end{center}

\section{The loop expansion and the mapping to the stochastic equation}
In the following a detailed derivation of the mapping discussed in the main text will be given. 
The derivation, albeit rather straightforward, requires basic knowledge of 
i) the field theoretical approach to second-order phase transition at a textbook level \cite{Parisi1988,LeBellac1991,Zinn-justin2002,Cardy1996}  ii) the $M$-layer expansion of \cite{altieri2017loop}  iii) the replica theory of the random field Ising model (RFIM) and the mapping to  stochastic equations leading to Parisi-Sourlas dimensional reduction \cite{parisi1992field}, an good compact presentation is given in sections 2.2-2.4 of \cite{de2006random}.
For the sake of readability some of the passages already discussed will be repeated. 

In the $M$-layer framework the average of a given observable over all the possible rewirings can be expressed as a sum over Feynman diagrams. 
A given Feynman diagram is transformed into a fat diagram substituting its  $I$ lines with one-dimensional chains of given lengths $\{L_1,\dots,L_I\}$ and attaching Bethe trees so that the connectivity of each vertex is exactly $c$. Within bootstrap percolation the natural observable is the joint probability that two or more sites are all on the $k$-core.  I will specialize to two-point correlations and thus to Feynman diagrams with two external vertexes. The derivation of the mapping can be easily generalized to higher orders correlations.

The computation of the occupation probability on a given fat diagram can be done it two steps: first we study the case in which all the vertexes of the corresponding Feynman diagram are occupied and perform eac on the internal lines and then we apply eac to the vertexes.
In order to do so we need to discuss the properties of a chain of length $L$ in the Bethe lattice. At variance with the single-site case, knowledge of the configuration of sites $s_1$ and $s_{L-1}$ prior to eac on sites $s_0$ and $s_L$ is not enough to perform eac on $s_0$ and $s_L$: one has to take into account the possibility that eac on $s_0$ ($s_L$) changes the value of $s_{L-1}$ ($s_1$). This happens if and only if site $s_0$ is emptied by eac and every site $\{s_1,\dots,s_{L-1}\}$ had exactly $m$ occupied neighbors before eac on $s_0$. Following standard terminology we say that sites $\{s_1,\dots,s_{L-1}\}$ belong to the corona cluster before eac on $s_0$ and $s_L$.
We introduce an additional variable $\chi$ that is equal to one if every site on the chain is on the corona and vanishes otherwise. If $\chi=0$  eac on $s_0$  ($s_L$) depends only on $s_1$ ($s_{L-1}$) before eac; if $\chi=1$  eac on $s_0$ and $s_L$ cannot be done independently but is easily performed: essentially  the whole line can be replaced by a single edge connecting site $s_0$ and $s_L$ directly.

Let us define  the difference between the probability distribution $P_L(s_1,s_{L-1},\chi)$ and its $L \rightarrow \infty$ limit  $P_B(s_1)P_B(s_L)\delta_{\chi,0}$:
\beq
\Delta P_L(\sigma,\tau,\chi)\equiv P_L(\sigma,\tau,\chi)-P_{B}(\sigma)P_{B}(\tau)\delta_{\chi,0} \ .
\eeq
In the next section we will derive the following large distance behavior:
\beq
\Delta P_L(\sigma,\tau,\chi)  =  \delta_{\chi,0}[g(\sigma)g(\tau)\, \Delta \,L\, \lambda^{L-1}+O(\lambda^L)]+
\delta_{\chi,1}\, \delta_{\sigma,1}\, \delta_{\tau,1}\lambda^{L-1}
\eeq
where $g(s)\equiv \delta_{s,1}-\delta_{s,0}$ and $\Delta$ is a positive constant. The function $g(s)$ has the property
\beq
\sum_s g(s)=0
\eeq
 that will be crucial in the derivation of the mapping.
Note that
the corona probability decays exactly as $\lambda^{L-1}$ but there is a larger $O(L \, \lambda^L)$ off-corona contribution.
The parameter $\lambda$ is computed in the next section, it has the following behavior close to the Bethe critical point $p=p_c$:
\beq
\lambda \approx {1 \over c-1}(1- \sigma^{1/2})
\label{lcrit}
\eeq
where $\sigma$ is a positive quantity vanishing linearly at the critical point  $\sigma \propto |p-p_c| $. 
As we will see in the following it is essential that $\lambda$ tends to $1/(c-1)$ at $p=p_c$ with a square root deviation.

The configuration of the vertexes of any Feynman diagram {\it after} eac over them is completely determined by: i) the configuration of the vertexes themselves after extraction but {\it before} culling, ii) the configuration of their neighbors  before culling on the Feynman vertexes but after eac on the lines iii) the values of the $\chi$'s of the lines.
Thus, for any Feynman diagram $G$ with two external vertexes we can define an indicator function $I_{12}$ that is equal to one if an only if site $s_1$ and $s_2$ (the external vertexes of the Feynman diagram) are both occupied after eac on all the vertexes:
\beq
I_{12}(G, s_1, \dots , s_{V'} ,\chi_1,\dots, \chi_{I})
\label{I12def}
\eeq
where  $V \geq 2$ is the total number of vertexes of $G$,  $V'=V+ c\, V$ is the number of sites including the $c$ sites at the frontier of each of each vertex and $I$ is the number of lines.
In order to obtain the probability that both sites $s_1$ and $s_2$ survive culling we have to sum the function $I_{12}$ over all possible configurations prior to culling with the corresponding probabilities.

It is convenient to specify for each edge $l$ of the Feynman diagram a direction, in field theory this is usually done to write the momentum conservation conditions at each  vertex and thus will be useful later, at this stage  it allows to define a function $in(l)$ ($out(l)$) that yields the index $j$ of the site at the beginning (at the end) of edge $l$.
We also define $S_e$ as the set of frontier sites that are on the edges of the Feynman diagram, {\it i.e.} the $2 I$ sites $s_sj$ such that $j=in(l),out(l)$ for $l=1,\dots,I$.
We then have:
\beq
\langle s_1 s_2 \rangle= \sum_{  s_1 ,\dots  s_{V'}, \chi_1,\dots, \chi_{I}  } I_{12} \left[\prod_{i=1}^V \left( p(s_i) \prod_{j \in \partial i, j \not \in S_e} P_{B}(s_j) \right)\right]\left[\prod_{l=1}^{I}P_{L_l}(s_{in(l) },s_{out(l)},\chi_l)\right]
\eeq
where the square brackets mean average over the initials configurations.
In the above formula $L_l$ is the length of edge $l$ and we have omitted the explicit dependence of $I_{12}$ from its arguments as given in (\ref{I12def}). 
Actually the correct objects to be computed in order the generate the $1/M$ expansion are the so-called line-connected observables for which a general expression was given in \cite{altieri2017loop}. This corresponds essentially to replace two-point correlations with {\it connected} two-point correlations and in this context  means that we have just to replace $P_L(\sigma,\tau,\chi)$ with $\Delta P_L(\sigma,\tau,\chi)$:
\beq
\langle s_1 s_2 \rangle_{lc}= \sum_{  s_1 ,\dots  s_{V'}, \chi_1,\dots, \chi_{I}  } I_{12} \left[\prod_{i=1}^V \left( p(s_i) \prod_{j \in \partial i, j \not \in S_e} P_{B}(s_j) \right)\right]\left[\prod_{l=1}^{I}\Delta P_{L_l}(s_{in(l) },s_{out(l)},\chi_l)\right] \ .
\eeq
The function $I_{12}$  depends only on the topology of the Feynman graph and not on the actual lengths of the lines. As discussed in \cite{altieri2017loop} in the critical region we must consider the large $L$ limit, therefore we should select the leading  $L \, \lambda^L$ off-corona contribution in $\Delta P_L$ for all lines. 
The key point is that such a contribution has a zero prefactor. Indeed if $\chi_l=0$ on all $l$ the function $I_{12}$ does not depend on all its arguments but only on the vertexes on the frontier of  $s_1$ and $s_2$. In particular it does {\it not} depend on any of the neighbors of all the other vertexes, as a consequence we can sum on any of the neighboring sites independently and this yields a term $\sum_sg(s)=0$ for each of them.
Thus when we select the leading contribution for a given line we must be careful not to get a vanishing contribution.

Let us concentrate on a single line $l$ for which we select that leading term $\chi_l=0$. 
We consider a couple of configurations of all the other variables (the occupancies of the sites and the $\chi$'s of the other edges) that differ only by the state (occupied or empty) of one of the sites of $l$, say $in(l)$. If the function $I_{12}$ does not change on these two configurations then again the term $\sum_{s_{in(l)}}g(s_{in(l)})=0$  leads to a vanishing contribution.
Thus a necessary condition for having a non-vanishing contribution is that the configuration of the other sites and of the other $\chi$'s must be  such that the internal vertex to which $s_{in(l)}$ is connected is on the same corona cluster of either $s_1$ or $s_2$ when $s_{in(l)}$ is occupied. This is the only case in which $I_{12}$  may change  when $s_{in(l)}$ is emptied.
Similarly if we assign the values of $\chi$ on each edge of the graph and study the effect of summing over the configurations of the sites on the edges with $\chi=0$, we can easily see that the necessary condition to have a non-vanishing contribution is that when all sites on those edges are occupied all sites connected to them must be on the same corona cluster of either $s_1$ or $s_2$.
The previous observations imply that in order to maximize the number of  off-corona edges to have the largest possible contribution while having a non-zero prefactor one has to arrange the corona edges on the graph on trees in such a way that each internal vertex is connected to either $s_1$ and $s_2$ by a path of corona edges. Once this condition is fulfilled we can set all other $\chi$'s  to zero thus selecting the leading contribution. 

The only configurations of the corona edges and of the vertexes that survive the sum over the off-corona edges vertexes are those such that all vertexes including $s_1$ and $s_2$ are on the corona when all sites on off-corona edges are occupied, so that $I_{12}=1$ on the latter and $I_{12}=0$ as soon as any of the off-corona edges is emptied.
Thus the total weight of those configurations is obtained multiplying for each vertex  a factor
\beq
p_{d_v}\equiv p {c-d_v \choose k-d_v} P^{k-d_v} (1-P)^{c-k}
\eeq   
 where $d_v$ is the connectivity (also called the degree) of the vertex in the Feynman diagram. In general in the $M$-layer context we have to consider Feynman diagrams with vertexes of degree up to $c$. Similarly to standard field theory one can argue that the leading diverging contribution at each order in the loop expansion are obtained considering  cubic Feynman diagrams with the lowest degree $d_v=3$ for all internal vertexes, unless they vanish for symmetry reason (which is not case here). For external vertexes we have simply $d_v=1$. 
More importantly we have a factor $\lambda^{L_l-1}$ and $\Delta {L_l}\, \lambda^{L_l-1}$ respectively for each corona and off-corona line.

According to \cite{altieri2017loop} the diagram must then be embedded on the original lattice (with $M=1$) assigning the positions of the vertexes and the incoming and outgoing directions of each line. For each line we have to multiply a factor $N_{L_l}(x,y,\mu,\mu')$  that yields the number of non-backtracking paths  of length $L_l$  starting at lattice point $x$ in the direction $\mu$ and ending at lattice point  $y$ from direction $\mu'$.
In the interesting regime  $L_l \gg 1$ one can argue that $N_{L_l}(x,y,\mu,\mu')$ is independent of the directions $\mu$ and $\mu'$ and can be replaced by  the total number of paths between $x$ and $y$ divided by the connectivity squared $N_{L_l}(x,y)/c^2$. Then the sum over different directions yields for each vertex $v$ a factor 
\beq
n_{d_v} \equiv d_v!{c \choose d_v} \ .
\eeq
It is  convenient to study the Fourier transform of the two-point function so that we have to sum over the positions of all  vertexes, including the external ones. In the relevant region where the vertexes are far apart the number of non-backtracking paths of length $L$ between two lattice points at $x$ and $y$ can be approximated by 
\beq
N_L(x,y)\approx {c (c-1)^{L-1} \over \rho} \, G(x-y)
\eeq   
where  $c(c-1)^{L-1}$ is the total number of paths starting from a given lattice point and $\rho$ is the density of the sites of the original lattice  (corresponding to $M=1$).  $G(x)$ is a Gaussian distribution with variance $2 \, D_{NBW} \, L$ where $D_{NBW}$ is by definition the diffusion coefficient of non-backtracking random walks in the original lattice.
Furthermore since the integrand varies slowly  we can replace the sum over the lattice points with an integral: $\sum_i \rightarrow \rho \int d^dx $.
The actual microscopic properties of the lattice encoded in $D_{NBW}$ and $\rho$ are irrelevant for the derivation of the mapping and indeed they will later be factorized. Performing also the internal integrals in Fourier space we end up with a momentum conservation condition (and a factor $(2 \pi)^{d-d \, d_v/2}$) at each internal vertex and a factor $\rho$ for each vertex.
Accordingly for each line we have a factor  
\beq
{1 \over c^2}N_L(k,k') \approx \delta(k+k')  { (c-1)^{L-1} \over c \, \rho}   \exp[- L D_{NBW} k^2]
\eeq 
where  the factor $c^2$ is to get the contribution at fixed incoming and outgoing directions.

At this point we perform the summation over $L$ on each line and we see that at the critical point since $\lambda$ obeys  eq. (\ref{lcrit}) the exponential decay is very slow and leads to a single pole contribution for a corona edge and a double pole contribution for an off-corona edge:
\beq
{1 \over c \, \rho}\left({1 \over \sigma^{1/2}+ D_{NBW} \, k^2 }\right)\ ,\  {1 \over c \,\rho} \Delta  \left({1 \over \sigma^{1/2}+ D_{NBW} \,k^2 }\right)^2 \ .
\label{bootpro}
\eeq 
Finally according to \cite{altieri2017loop} each diagram must be divided by the symmetry factor of the Feynman diagram $S(G)$ and multiplied by a factor $1/ M^{n+ L-1}$ where $n$ is the order of the observable (the number of external vertexes of the diagram) and $L$ is the number of loops.
If we rescale the distances in order to have $D_{NBW} k^2 \rightarrow k^2$ we have additional factors for each internal line and for each internal vertex.
We have thus constant factors $p^{in}$ associated to internal vertexes, $p_{ext}$ to external vertexes, $i_{in}$ to internal lines and $i_{ext}$ to external lines. 
For cubic diagrams internal vertexes and internal lines have expressions $V_{in}=2(L-1)+n$ and  $I_{in}=3(L-1)+n$
thus if we also redefine the observable of order $n$ multiplying it by an appropriate constant $b^n$  all factors (except the $(2 \pi)^{d-d \, 3/2}$ factor at each internal vertex) can be  reabsorbed into a rescaling of the factor $M$:
\beq
{1 \over \, M^{n+L-1}}b^n i_{in}^{I_{in}} i_{ext}^n p_{in}^{V_{in}} c_{ext}^n p_{ext}^n = {1 \over \tilde{M}^{n+L-1}}\ .
\eeq
Thus we arrive at the final result that the (rescaled) observable of order $n$ is obtained summing over all cubic Feynman diagrams with a factor $1/(\tilde{M}^{n+L-1}S(G))$. For each diagram we have to consider all possible valid arrangements of corona and off-corona edges and multiply the propagators (\ref{bootpro}) without the $(c \rho)^{-1}$ factors and with $D_{NBW}=1$. Then we have to perform the integrals over the internal momenta with the momentum conservation condition and a factor  $(2 \pi)^{d-d \, 3/2}$ at each internal vertex.

We will now show that the above Feynman rules are exactly the same of a spinodal in a random field. Let us recall the definition of the corresponding replicated action in terms of fields $m_a(x)$ where   $a=1,\dots n$ and $n \rightarrow 0$ at the end:
\beq
\int \left(\prod_{a,x} d m_a (x)\right)\, \exp \left[ -  \tilde{M} \mathcal{L}(m) \right]
\eeq
\beq
\mathcal{L}(m) \equiv \sum_{a=1}^n \,\int d^dx\,\left( {\sigma \over 2} m_a(x)+{1 \over 2} |\nabla m_a(x)|^2  -{1 \over 3!} m_a^3(x) \right)  - {\Delta \over 2 } \int d^d x \, \left( \sum_a m_a(x)\right)^2
\label{repaction}
\eeq
The loop expansion around the stable mean-field solution $m_a(x)=-\sigma^{1/2}$ is controlled by the propagator:
\beq
\langle m_a(k) m_b(k') \rangle_c = \delta(k+k') \left[  {\delta_{ab} \over \sigma^{1/2}+ k^2 }+ {\Delta } \left({1 \over  \sigma^{1/2}+ k^2 }\right)^2   \right]
\eeq
Note that due to the $n\rightarrow 0$ limit simple scaling badly fails and the propagator displays a double pole, following the literature we associate a continuous line to the single-pole term and a crossed line to the double-pole term. Given a Feynman diagram we would like  to select the most diverging contribution but care must be taken because of the replica summations.
If we select the crossed term on each line entering a given vertex the summation over  the  vertex replica indexes yields $\sum_a=n=0$. A standard analysis, see {\it e.g.} \cite{de2006random} sec. 2.2-2.4, tells us that the leading non-vanishing contributions are obtained arranging the simple and crossed propagators in such a way that all vertexes are on trees of simple propagators connected to the external vertexes.
Thus we see that the Feynman rules for the most diverging diagrams are exactly the same that we have obtained above for bootstrap percolation. Furthermore the factor $(2 \pi)^{d-d \, 3/2}$ for the internal vertexes appears when we write the cubic vertex contribution in Fourier space.
One can also show that the most diverging diagrams are those generated by the solution of a stochastic equation. Through a Hubbard-Stratonovich transformation we rewrite the term proportional to $\Delta$ in (\ref{repaction}) as:
\beq
 \int \left[\prod dh(x)\right] e^{-\int d^dx {h^2(x) \over 2 \Delta}+\tilde{M} \int d^d x \, {h(x) \over \tilde{M}^{1/2}}  \sum_a m_a(x)} \ .
\eeq
Different replicas are now decoupled and a loop expansion of each replica separately before the integration over the random fields generates all possible diagrams with simple poles and sources $h(x)$. The subsequent average over the Gaussian random fields generates the double poles through the merging of two sources $h(x)$. 
Since, as we saw above, the relevant diagrams {\it after} the average over the random sources are those where the simple poles are arranged on trees we can select them {\it before} the average. A classic field-theoretic result states that tree diagrams are generated by the solution of the mean-field equation:
\beq
{\sigma \over 2 }  +{h(x) \over \tilde{M}}- \nabla^2 m(x)- {1 \over 2}m^2(x)=0
\eeq
and therefore we can write for an $N$-point function in the replicated system
\beq
\langle m_{a_1}(x_1)\dots m_{a_N}(x_N) \rangle \approx [m(x_1)\dots m(x_N)]
\eeq
where $m(x)$ is the solution of the above stochastic equation and the square brackets mean average over the solution of the above stochastic equation with Gaussian random fields $h(x)$
\beq
[h(x)]=0\, , \ \ [h(x)h(y)]= {\Delta} \, \delta(x-y)
\eeq
 Note that in principle the averages over the random fields have a weight $Z^n$ and are not white, but they become white averages at $n=0$.

\section{Two-point correlations on the Bethe lattice}

In the following we will derive the expression of the two-point function through an iterative equation. We start noticing that the probability that every site of a chain of length $L$ is on the corona cluster is given exactly by $\lambda^{L-1}$ where the exact expression of $\lambda$ is
\beq
\lambda=p {c-2 \choose k-2 }P^{k-2}(1-P)^{c-k} 
\eeq
Therefore we have $\sum_{s_1,s_{L-1}}P_L(s_1,s_{L-1},0)=1-\lambda^{L-1}$. 
Next we write an iterative equation for $P_{L+1}(s_L,s_1,\chi)$ in terms of $P_L(s_{L-1},s_1,\chi)$ performing eac on $s_L$ while keeping $s_{L+1}$ occupied.
Unless every site $\{s_1,\dots, s_{L-1}\}$ was on the corona, the process of eac of $s_{L}$ will not affect $s_1$; this case will contribute a term 
\beq
\left(
\begin{array}{c}
P_{L+1}(1,s_1,0)
\\
P_{L+1}(0,s_1,0 )
\end{array}
\right) =
\left(
\begin{array}{cc}
p\, P_{k-2,c-2} & p\, P_{k-1,c-2}
\\
1-p\,   P_{k-2,c-2} &  1-p \, P_{k-1,c-2}
\end{array}
\right)\,
\left(
\begin{array}{c}
P_{L}(1,s_1,0)
\\
P_{L}(0,s_1,0 )
\end{array}
\right)+\dots
\eeq
where the dots represent the corona contribution to be discussed later.
The above matrix has two eigenvalues, the Perron-Frobenius (PF) eigenvalue $\lambda_{PF}=1$ with right eigenvector $(P,1-P)$ and left eigenvector $(1,1)$ and another eigenvalue equal to $\lambda$ controlling the corona probability with right eigenvector $(1,-1)$  (orthogonal to left PF eigenvector $(1,1)$).

The additional contribution to $P_{L+1}(s_L,s_0,0)$ comes from those configurations that were on the corona prior to eac on $s_{L}$ but are no longer on the corona after. This can either happen, after eac, if  site $s_{L}$   is still occupied with more than $m$ occupied neighbors or if it is emptied. The first case contributes to $P_{L+1}(1,1,0)$, the second to $P_L(0,0,0)$ and the sum of the two contributions must be equal to the probability that a corona configuration is no longer a corona configuration after culling {\it i.e.} $\lambda^{L-1}(1-\lambda)$.

Now let us write $P_L(s_{L-1},s_L,0)$ in full generality in terms of the eigenvectors of the matrix, we will have:
\beq
P_L(\sigma,\tau,0)  =  a_{11} ^{(L)}P_{B}(\sigma)P_{B}(\tau)+ a_{12}^{(L)}P_{B}(\sigma)g(\tau)+  a_{21}^{(L)}g(\sigma)P_{B}(\tau)+ a_{22}^{(L)} g(\sigma)g(\tau)
\eeq
The normalization condition implies $a_{11}^{(L)}=1-\lambda^{L-1}$ while symmetry of the function with respect to exchange of its arguments implies 
$a_{12}^{(L)}=a_{21}^{(L)}$.
The iterative equation translates then into an equation for the components $a_{ij}^{(L)}$. Since the application of the linear matrix breaks the symmetry with respect to the exchange while all other terms in the equation are symmetric we must have $a_{12}^{(L)}=a_{21}^{(L)}=0$.
For the same reason the contribution coming from the corona cannot have a mixed term of the form $P_B(\sigma)g(\tau)+P_B(\tau)g(\sigma)$, besides since it is concentrated  on $(1,1)$ and $(0,0)$ and zero otherwise it must be equal to:
\beq
\lambda^{L-1}(1-\lambda) (P_{B}(\sigma)P_{B}(\tau)+P (1-P) g(\sigma) g(\tau))\ 
\eeq
This leads to the following equation for $a_{22}$
\beq
a_{22}^{(L+1)}=\, \lambda \, a_{22}^{(L)}+\lambda^{L}\Delta\ ,
\eeq
where 
\beq
\Delta \equiv {1-\lambda \over \lambda} P(1-P) \ 
\eeq
The above equation is valid for $L\geq 2$ with $a_{22}^{(2)}=\lambda \Delta$
and the exact solution is then 
\beq
a_{22}^{(L)}= (L-1) \,\lambda^{L-1}\, \Delta 
\eeq
leading to the final expression for the propagator:
\beq
\Delta P_L(\sigma,\tau,\chi)  =  \delta_{\chi,0}[g(\sigma)g(\tau)\, \Delta \,L\, \lambda^{L-1}+O(\lambda^L)]+
\delta_{\chi,1}\, \delta_{\sigma,1}\, \delta_{\tau,1}\lambda^{L-1}\ .
\eeq
Performing eac on the extrema $s_0$ and $s_L$ we may obtain the correlation on the Bethe lattice at leading order: 
\beq
\langle s_0 s_L\rangle=P_{site}^2+{p_1^2\, \Delta \over \lambda}\,L\, \lambda^{L}+O(\lambda^L)\ 
\eeq
where:
\beq
p_1 \equiv p {c-1 \choose k-1} P^{k-1} (1-P)^{c-k}\ 
\eeq

\end{document}